%% file: samplepaper.tex
\documentclass[runningheads]{llncs}
\usepackage[T1]{fontenc}
\usepackage{graphicx}
\usepackage{amsmath}
\usepackage{booktabs}
\usepackage[protrusion=true,expansion=false]{microtype}
\begin{document}
\title{LoFi RADIO: A Distilled In-Domain Backbone Applied for Artifact-Severity\\
Grading of Ultra-Low-Field Neonatal Brain MRI}
\titlerunning{Distilled In-Domain Backbone for ULF Neonatal MRI QC}
% MUST BE ANONYMIZED FOR INITIAL SUBMISSION
\author{Jonathan B. Martin\inst{1}\orcidID{0000-0002-9384-8056},
Yashwant Kurmi\inst{1}\orcidID{0000-0003-4986-2106}, \and
Charlotte R. Sappo\inst{1}\orcidID{0000-0002-7030-278X}}
% %
\authorrunning{J.B. Martin et al.}
% %
\institute{$^1$Vanderbilt University Institute of Imaging Science, Nashville, TN, USA
\email{jonathan.bach.martin@vumc.org}}
\maketitle              % typeset the header of the contribution
\begin{abstract}
Ultra-low-field MRI makes neonatal brain imaging deployable in low-resource
settings, but its low SNR, lack of shielding, and long scan duration make it especially prone to acquisition artifacts,
motivating automated quality control. We address the LISA~2026 Task~1a challenge: multi-label severity
grading (0/1/2) of seven common image artifacts on ULF $T_2$ weighted volumes. We identify that a number of backbones may be successfully paired with a classification MLP, but that no single backbone is uniformly best across artifacts. To improve performance, we evaluate 
routing complementary foundation model teachers through a per-artifact gate, as well as distilling the teachers into a single in-domain ViT-S student (LoFi RADIO) over an unlabeled low-field MRI corpus. Both of these strategies improve the weighted composite. The distilled backbone matches or exceeds the gate and has the added advantage of not requiring deployment of multiple large foundation models at inference. 

\end{abstract}
\section{Introduction}
% TODO: expand. Stub kept brief; this draft focuses on Methods.
Ultra-low-field (ULF) MRI at $0.064$\,T offers a low-cost, portable option for
neonatal neuroimaging in resource-limited settings~\cite{deoni2021lowfield}. Despite its advantages, successful ULF MRI is challenged by intinsically low
signal-to-noise ratio, lack of shielding, and long acquisition times. Imaging of neonates introduces even greater potential for image quality challenges, as neonates frequently move during scans and are often imaged using RF coils designed for adults, exacerbating image quality issues. Together, these challenges make ULF scans of neonates particularly
susceptible to a variety of artifacts. Furthermore, many commercial low field systems may be operated by users with minimal experience, not requiring ARRT MR technologist certification. Automated, per-volume quality control (QC) is therefore
required for reliable use of artifact-susceptible systems with minimal expert supervision. The LISA (\textbf{L}ow field pediatric brain magnetic resonance \textbf{I}mage \textbf{S}egmentation and quality \textbf{A}ssurance Challenge)~2026 Task~1a
benchmark~\cite{lisa2026} provides a setting to evaluate approaches to the relevant multi-label artifact severity grading QC task. This is a challenging classification task, requiring generalist performance across images sourced from multiple low field scanners with distinct image quality and artifacts. 

Foundational Models (FMs), large multi-purpose models trained on large amounts of often unlabeled data, have emerged as highly effective backbones capable of excelling at a wide range of vision and image processing tasks. It has been demonstrated that multiple FMs trained for complementary domains can outperform a single FM when combined into a single student model via multi-teacher knowledge distillation~\cite{ranzinger2024amradio}. Distillation may be performed with the objective of generating a model for a more specialized set of tasks, for example by distilling on a set of data which includes examples from the target domain~\cite{Hinton2015}. Distillation also provides computational advantages, as larger models may be distilled into a single, more compact student. Our approach adapts the method of Agglomerative Vision Foundation Model -- Reduce All Domains Into One (AM-RADIO)~\cite{ranzinger2024amradio} but includes specialist student distillation on a dataset of low-field MRI images, with the objective of developing a distilled foundational model optimized for low-field MRI vision tasks. Thus we call our approach Low-Field RADIO (LoFi RADIO). LoFi RADIO produces a more hardware-efficient backbone than strategies involving multiple distinct foundation models. It was identified as having the highest overall composite performance on the LISA Task1a metrics of all single and combined FM approaches that were evaluated in this study, as well as 3rd-best mean composite performance out of 16 submitting teams on a blinded LISA Task 1a validation dataset. 

\begin{figure}[t]
\includegraphics[width=\textwidth]{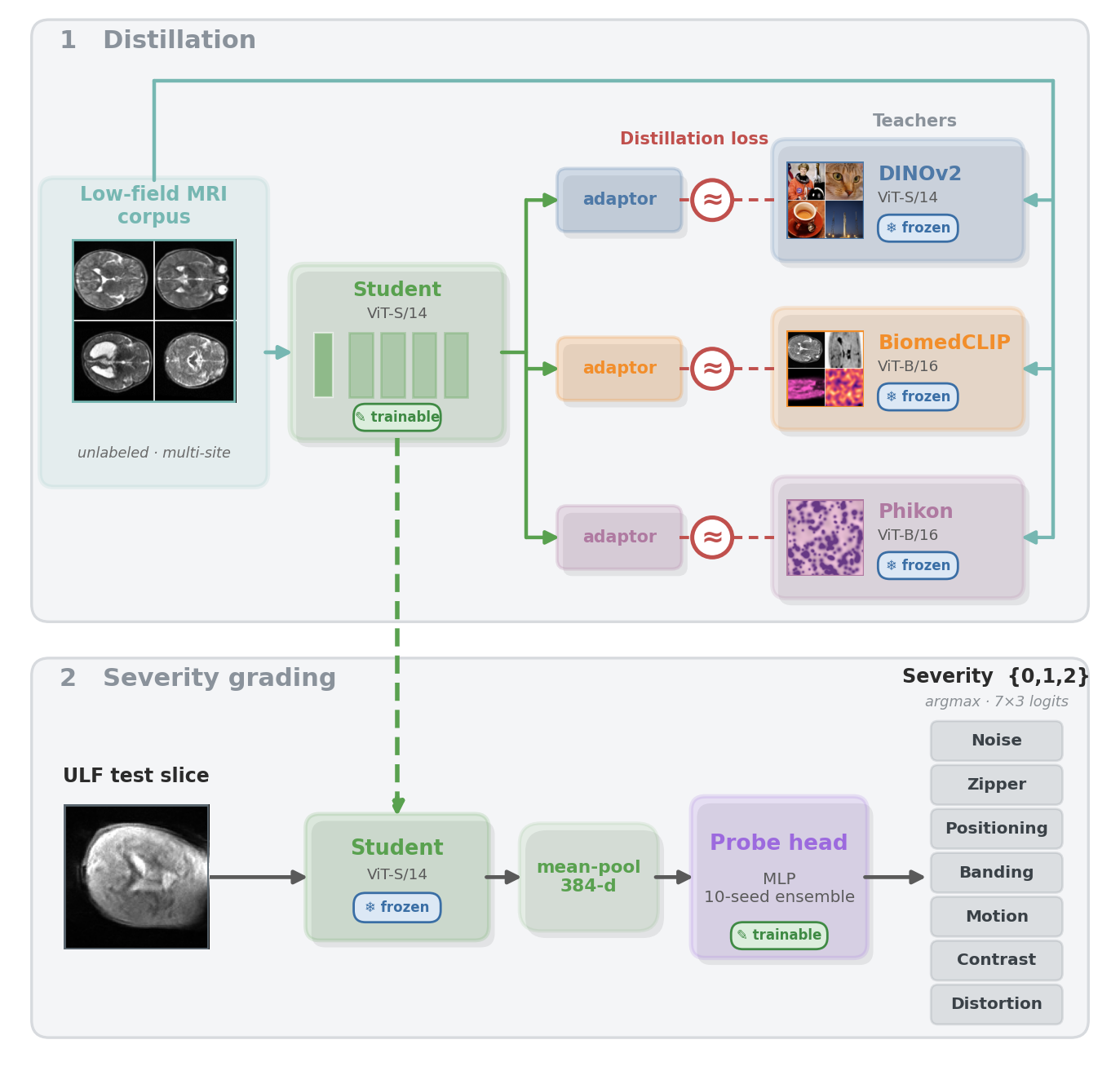}
\caption{Overview of the LoFi-RADIO pipeline. \textbf{(Stage 1)~Distillation:} the same
unlabeled low-field slice is encoded in parallel by frozen foundation-model teachers
(DINOv2, BiomedCLIP, Phikon) and by a trainable ViT-S/14 student; per-teacher adaptor
heads map the student's features into each teacher's space. The
teachers are frozen and only teach the student. \textbf{(Stage 2)~Artifact severity grading:}
the trained student is frozen and mean-pooled to a $384$-d descriptor, on which a
tuned MLP probe predicts a $\{0,1,2\}$ severity grade for each of the seven artifacts.
} \label{fig0}
\end{figure}

\section{Methods}
\label{sec:methods}

\subsection{Task, Data, and Evaluation Metric}
The LISA 1a 2026 challenge includes images with seven common artifacts---'Noise', 'Zipper', 'Positioning', 'Banding', 'Motion', 'Contrast',
and 'Distortion'---each on an ordinal severity scale of $0$ (none), $1$ (moderate),
$2$ (severe), per volume. The data are reconstructed, magnitude-only $T_{2}$ weighted
NIfTI volumes (531 single-orientation acquisitions from 243 subjects imaged
in axial, coronal, and sagittal planes; each orientation is labeled independently)
from three sites. Because the data are magnitude-only, the phase information that
encodes motion- and zipper-type corruption is discarded at reconstruction and cannot be leveraged for this task.

Each volume is reduced to  slices whose above-background area exceeds a fixed
fraction, using an intensity threshold at 10\% of the volumes maximum intensity. Every
retained slice is then intensity-normalized independently,
taken between a
median${}\pm{}k\!\cdot\!$MAD window computed over the in-slice foreground to remove scan and site intensity variations.

The official LISA competition metric averages five support-weighted classification metrics
(accuracy, F1, F2, precision, and recall) computed over the $\{0,1,2\}$ grades into a per-artifact
composite. The headline composite score is the mean of this composite over the seven artifacts. Because
the metric is support-weighted and the grade-$0$ class dominates (most scans are artifact-free), we used a balanced
cross-entropy objective with $\arg\max$ decoding. Unless otherwise noted, differences in composite score were assessed against a
multi-seed noise floor of $\pm0.0064$, corresponding to two standard deviations.

\subsection{Frozen Foundation-Model Probe}
\label{sec:probe}
Our base classification model uses a frozen self-supervised vision backbone as a fixed
feature extractor. We applied the following vision models as frozen backbones: DINOv2~\cite{oquab2024dinov2}, SAM,~\cite{kirillov2023sam}, SAM2~\cite{ravi2024sam2}, BrainIAC,~\cite{brainiac}, BiomedCLIP~\cite{zhang2023biomedclip}, Phikon~\cite{filiot2023phikon}, MAE~\cite{he2022mae}, RadioDINO~\cite{zedda2025radiodino}, Triad~\cite{wang2025triad}, and ARNIQA~\cite{agnolucci2024arniqa}. Table 1 outlines these models and some of their characteristics. Each 2D model was applied in a 2.5D fashion, in which each slice in the volume was encoded independently and the resulting per-slice summary
(\texttt{CLS}) embeddings were mean-pooled into a single $384$-dimensional descriptor for the
volume. The two 3D models (BrainIAC and Triad) were run natively in 3D. A small multilayer perceptron head mapped the descriptor output to artifact severity logits
and was trained with a cross-entropy loss. Only the head and its optimizer were tuned, using a random hyperparameter
search over a subject-grouped five-fold objective. The best-performing configuration, consisting of a
single hidden layer with a dropout of $0.5$ and a learning rate of $5\times10^{-4}$, and was used in all following experiments. 

Because the relative utility of a candidate backbone could not be reliably predicted in
advance, we evaluated each candidate directly. We evaluated every candidate using a single-arm probe under one identical protocol,
with the same tuned head, folds, and seeds used elsewhere in this work, and reported each result
paired against DINOv2, considered the reference vision model. We also assessed two simple degeneracy probes, meant to aid in screening out models that map nearly every volume to nearly-identical feature outputs. These were based on the effective rank~\cite{roy2007effrank} ('eff-rk' in Table~\ref{tab:screening}) and mean off-diagonal cosine similarity across volumes ('$\overline{|\cos|}$' in Table~\ref{tab:screening}) .  Results compiled in Table~\ref{tab:screening} informed selection of frozen backbones used in combination methods.

\input{tab_screening}

\subsection{Per-Artifact Backbone Gate}
\label{sec:gate}
It was hypothesized that no single frozen backbone would be uniformly best across all seven artifacts, due to their distinct objectives and training domains. To exploit their complementary strengths, we explored two simple multi-foundation model approaches prior to evaluating a more complex distillation approach: 1) a simple concatenation of output features from models, and 2)
 a per-artifact gate over four teachers: DINOv2, BiomedCLIP,
SAM, and
Phikon. These models were selected for their relative non-degeneracy as measured by eff-rk compared to other models (Table~\ref{tab:screening}). DINOv2 was chosen over RadioDINO due to its superior composite score, despite inferior eff-rk. Each backbone $b$ produced an independently z-scored feature vector and its own probe
head, and for each artifact $a$ a learned softmax weight $\alpha_{a,b}$ routed and ensembled the
per-backbone predictions,
\begin{equation}
\hat{p}_a \;=\; \sum_{b} \alpha_{a,b}\, p_{a}^{(b)},
\qquad \sum_b \alpha_{a,b}=1 ,
\end{equation}
with the routing weights fit under the same leak-free protocol used throughout.

\subsection{LoFi-RADIO: Agglomerative In-Domain Model Distillation}
\label{sec:radio}
While the gate may potentially improve performance, it requires multiple backbones and a router at inference time, and
its features remain native to the out-of-domain data on which the teachers
were pretrained. To address both limitations, we distilled the complementary teachers into a
single in-domain student backbone, which we refer to as LoFi-RADIO, following the approach of
agglomerative foundation-model distillation~\cite{ranzinger2024amradio}. The student was a ViT-S/14 initialized from DINOv2, and was trained to reproduce the
representations of DINOv2, BiomedCLIP, and Phikon. The resulting $384$-dimensional student
mean-pool serves as a direct replacement for the descriptor consumed by the probe of
Section~\ref{sec:probe}.

\paragraph{Distillation objective.}
For each teacher $t$, following AM-RADIO, we attached a lightweight adaptor MLP $g_t$ (2 layers, LayerNorm and GELU activation) that mapped the student
embedding into that teacher's feature space. The cached teacher targets were
standardized on-device for each batch using a PHI-S-style standardization~\cite{ranzinger2024phis},
which places the heterogeneous teachers on a common scale. We supervised both the summary token
and the dense ($16\times16$) spatial features,
\begin{equation}
\mathcal{L} \;=\;
\sum_{t}\Big[\, \underbrace{1-\cos\!\big(g_t(s_{\text{sum}}),\, \tilde{z}^{t}_{\text{sum}}\big)}_{\text{summary cosine}}
\;+\;
\underbrace{\big(1-\cos(\cdot)\big) + \mathrm{SmoothL1}(\cdot)}_{\text{dense, per location}}\,\Big],
\end{equation}
where $\tilde{z}^{t}$ denotes the standardized teacher target. We observed that not every gate
member transferred usefully to distillation. The frozen SAM feature is low-rank and
contributes to the gate only through routing, and because a single fused student cannot reproduce
a routed signal, distillation of SAM was limited to the performance of feature concatenation. We
therefore excluded SAM from the student's teacher set. 

Because distillation is unsupervised and requires only images, we were able to pool labeled and
unlabeled data freely. The corpus combined the LISA training images with external low-field brain
MRI, comprising M4Raw~\cite{lyu2023m4raw} ($0.3$\,T adult) and two exact-field $0.064$\,T adult T2
datasets~\cite{lf64mt,ds006557}, loaded through a unified corpus interface that standardized
loading and canonicalization across sources. Distillation was parallelized across
multiple GPUs using synchronous data parallelism, since the pooled dense target cache was too
large to replicate per device, with the student and all adaptors held in a single module so that
the per-GPU losses were gathered correctly.
\begin{figure}[t]
\includegraphics[width=\textwidth]{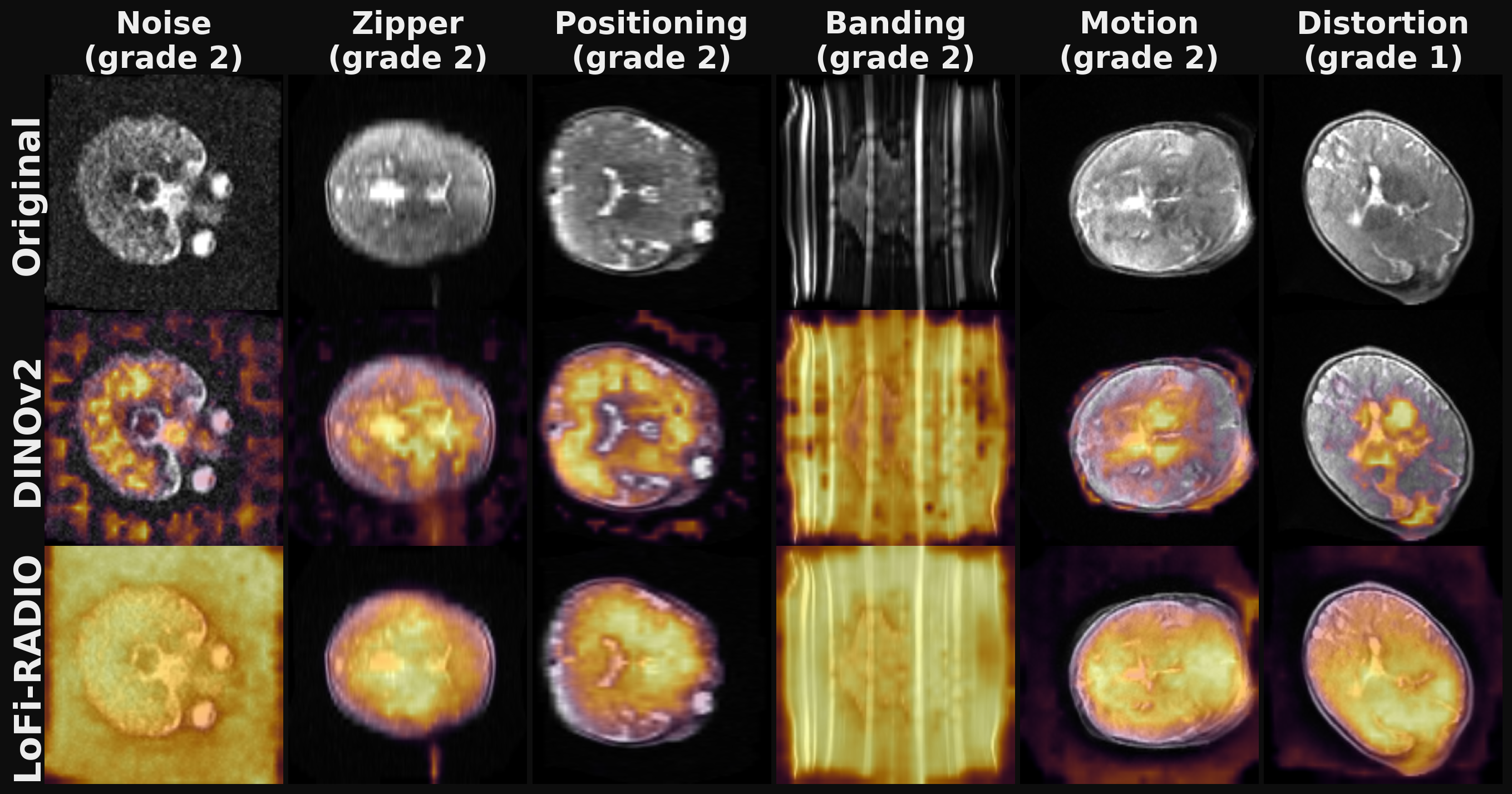}
\caption{Per-artifact discriminative attention for the LoFi-RADIO student. Columns are
  representative high-severity volumes for different artifacts (grade~2, except Distortion,
  shown at grade~1); rows are the input slice and the attention of the distilled student and of
  DINOv2. We project each patch token onto the grade-2-minus-grade-0 mean-feature direction
  $\alpha$ (warmer = toward the severe prototype); since the probe pools patch tokens, this is the
  exact spatial decomposition of the near-linear severity readout. For several artifacts the
  response aligns with the expected physical signature (e.g.\ Zipper with the background stripe,
  Distortion with greater values towards the $B_0$-perturbed regions , Motion with marginal ghosting), whereas for others it
  is more diffuse, spread across much of the brain---consistent with the global
  nature of those corruptions.
} \label{fig1}
\end{figure}

\subsection{Leak-Free Evaluation of the Distilled Student}
\label{sec:leakfree}
Because the LISA images are present, without labels, in the distillation corpus, distilling a
single student on all images and then cross-validating the probe would introduce a subtle form of
leakage, since the student would have seen the validation-fold images during distillation, an
advantage that the frozen baselines never receive. To avoid this, we adopted a leak-free per-fold
protocol. For each of the five subject-grouped folds, we distilled a separate student on a corpus
that excluded that fold's validation images while still including the remaining LISA training
images and all external data, and we predicted each fold using only the student that had never
seen it. Assembling these held-out predictions yields an out-of-fold composite that is directly
comparable to the frozen DINOv2 and gate baselines. For the final challenge submission, in which
the test set is disjoint and withheld by the organizers, no such exclusion is required. In this
case we distilled a single student on all available data, exported its pooled LISA features,
trained an ensemble of tuned probe heads on all labeled LISA volumes, and applied the resulting
model to the test volumes using the identical 2.5D pooling procedure to produce per-volume
$\{0,1,2\}$ grades.

% \subsection{Implementation Details}
% \label{sec:impl}
% Models are implemented in PyTorch. Backbones are applied 2.5D with foreground-slice
% selection and slice-wise mean-pooling. Probe heads use cross-entropy with cosine
% learning-rate annealing and AdamW; the submission probe averages the softmax outputs
% of a $10$-head seed ensemble before $\arg\max$ decoding. All comparisons between a
% candidate and a baseline are made at matched seeds and judged paired, against the
% $\pm0.0064$ composite noise floor.

% Method ablation: Single FM / Concat / Gate / Distilled student.
% Generated by paper/tab_method_ablation.py from tri_probe JSONs (2026-06-26). Needs \usepackage{booktabs}.
% Source: matched 10-seed subject-grouped CV, tuned probe head (dropout 0.5, lr 5e-4),
%         official support-weighted composite. Keep in sync via: python paper/tab_method_ablation.py
\begin{table}[t]
\centering
\caption{Comparison of single- and multi-backbone strategies for ULF neonatal artifact grading, evaluated using the matched 10-seed, subject-grouped five-fold protocol. Per-artifact and overall composite scores are reported as mean$\pm$standard deviation, with the best value in each column shown in \textbf{bold}. A single distilled student, using one ViT-S/14 backbone, exceeded the four-backbone concatenation and the per-artifact gate without requiring multiple backbones or a router. The concatenation did not improve upon the single backbones under the support-weighted metric, while the gate recovered part of the complementarity through routing. The distilled configurations are shown cumulatively, adding the Phikon teacher and then the exact-field OpenNeuro corpus to the DINOv2 and BiomedCLIP student.}
\label{tab:method_ablation}
\setlength{\tabcolsep}{4.5pt}
\small
\resizebox{\textwidth}{!}{%
\begin{tabular}{lcccccccc}
\toprule
Method & Noise & Zip. & Pos. & Band. & Mot. & Con. & Dist. & Composite \\
\midrule
\multicolumn{9}{l}{\textit{Single frozen FM}} \\
DINOv2 & 0.909 & 0.810 & 0.855 & \textbf{0.973} & 0.749 & 0.756 & 0.724 & 0.8249\,$\pm$\,0.0022 \\
BiomedCLIP & 0.901 & 0.818 & \textbf{0.867} & 0.966 & 0.732 & 0.757 & 0.734 & 0.8240\,$\pm$\,0.0041 \\
SAM & 0.911 & 0.800 & 0.858 & 0.958 & 0.716 & 0.750 & 0.711 & 0.8158\,$\pm$\,0.0035 \\
Phikon & 0.913 & 0.804 & 0.864 & 0.964 & 0.745 & 0.772 & \textbf{0.738} & 0.8287\,$\pm$\,0.0028 \\
\midrule
\multicolumn{9}{l}{\textit{Combination of all four FMs}} \\
Concat & 0.913 & 0.823 & 0.861 & 0.968 & 0.743 & 0.767 & 0.730 & 0.8293\,$\pm$\,0.0037 \\
Gate (per-artifact router) & \textbf{0.914} & 0.821 & 0.866 & 0.966 & 0.753 & 0.771 & \textbf{0.738} & 0.8327\,$\pm$\,0.0030 \\
\midrule
\multicolumn{9}{l}{\textit{Distilled student (ours)}} \\
Student~$\leftarrow$~D+BMC & 0.912 & \textbf{0.832} & 0.863 & 0.970 & 0.756 & 0.765 & 0.731 & 0.8330\,$\pm$\,0.0038 \\
\quad +~Phikon (3 teacher) & \textbf{0.914} & 0.823 & \textbf{0.867} & 0.967 & 0.760 & 0.777 & 0.736 & 0.8349\,$\pm$\,0.0030 \\
\quad +OpenNeuro corpus & 0.912 & 0.830 & 0.865 & 0.967 & \textbf{0.769} & \textbf{0.781} & \textbf{0.738} & \textbf{0.8374\,$\pm$\,0.0025} \\
\bottomrule
\end{tabular}%
}
\end{table}

% \begin{figure}
% \includegraphics[width=\textwidth]{fig1.eps}
% \caption{LoFi-RADIO pipeline: frozen multi-teacher gate (top) distilled into a
% single in-domain ViT-S student over an exact-field 0.064\,T corpus (bottom).}
% \label{fig:pipeline}
% \end{figure}

\section{Results}
\label{sec:results}
All comparisons were made using the matched $10$-seed, subject-grouped
five-fold protocol with the tuned probe head, and were evaluated using the official
support-weighted composite. Differences were assessed against the multi-seed noise floor of
$\pm0.0064$ described above.

\subsection{Complementary teachers and the per-artifact gate}
\label{sec:res_gate}
When each backbone was screened using the same single-arm probe (Table~\ref{tab:screening}),
frozen utility varied little across pretraining objectives: every candidate but ARNIQA fell within
approximately $0.013$ of DINOv2, and neither degeneracy metric was predictive of the result. Among the four gate members, the single frozen backbones scored $0.8249$ (DINOv2),
$0.8240$ (BiomedCLIP), $0.8158$ (SAM), and $0.8287$ (Phikon). No single backbone was uniformly
best, and each captured a subset of the rare classes (Table~\ref{tab:method_ablation}).
A naive concatenation of all four backbones did not convert this complementarity into an
improvement, reaching $0.8293$. The per-artifact gate, which routed and ensembled the per-backbone predictions, recovered
performance, reaching $0.8327$ for the four-backbone gate. 
% The learned routing weights concentrated on the backbone that
% probed best for each artifact rather than averaging uniformly, and the resulting gains were
% largest on the weak, rare classes (Zipper, Distortion, Positioning, and Motion) that motivated the
% gate.

\subsection{A single distilled backbone matches and exceeds the gate}
\label{sec:res_distill}
Distilling the complementary teachers into a single in-domain ViT-S student removed the
requirement for multiple backbones and a router at inference, while retaining and ultimately
exceeding the accuracy of the gate (Table~\ref{tab:method_ablation}). A student distilled from only
DINOv2 and BiomedCLIP reached a composite of $0.8330$, performing similarly to the four-backbone gate ($0.8327$). Adding the texture-sensitive Phikon teacher increased this to $0.8349$. Pooling the exact-field
$0.064$\,T OpenNeuro corpus into the distillation corpus further increased the composite
to $0.8374$. The per-artifact gains relative to DINOv2 were primarily within the classes which
the gate had provided its largest improvements, namely Contrast, Distortion, and Motion
(Table~\ref{tab:method_ablation}). These results suggest that the in-domain distillation process, and in
particular the Phikon teacher, was the dominant driver of the improvement over the gate.

\subsection{Contribution of the in-domain distillation corpus}
\label{sec:res_corpus}
The overall composite was largely saturated by distillation on the LISA images alone ($0.8349$).
Adding the external low-field corpus left the overall composite unchanged within noise ($0.8374$)
providing evidence only that the improvement was attributable to the distillation and the Phikon teacher rather
than to the external corpus itself. What the exact-field corpus changed was the distribution of
this improvement across artifacts, redistributing model capacity toward the weak
classes and reducing the across-seed variance.

\subsection{LISA official validation performance}
Table~\ref{tab:validation} shows the  individual weighted metric and composite scores for the official LISA validation dataset. These results were from an additional collection of images with ground-truth IQA scores withheld from LISA challenge participants. On the held-out dataset, LoFi RADIO reached a best composite score of 0.8322, 3rd of 16 submitting teams (Team \#1 Best Composite: 0.8356, Team \#2 Best Composite: 0.8334, Team \#3 (LoFi RADIO) Best Composite: 0.8322).

\begin{table}[t]
\centering
\caption{Official held-out validation results for the final submission (top-ranked entry).
The challenge metric averages five support-weighted scores over the $0/1/2$ severity grades.}
\label{tab:validation}
\begin{tabular}{lccccc c}
\toprule
& Accuracy & Precision & Recall & F1 & F2 & Composite \\
\midrule
LoFi-RADIO & 0.842 & 0.833 & 0.842 & 0.816 & 0.828 & \textbf{0.8322} \\
\bottomrule
\end{tabular}
\end{table}

\section{Discussion}
In this study, we examined whether a single compact foundation-model backbone, distilled in-domain from several complementary
teachers, can outperform single- or multi-foundation model backbones on an ULF MRI image quality assessment task. We observed that a frozen foundation model in most cases serves as a strong baseline feature
extractor for this task. However, no single frozen backbone is best across all artifacts. Artifact-specific routing through a trained gate recovers complementary model strengths better than simple concatenation. However
multi-teacher distillation can consolidates those strengths into one in-domain student with further improved performance.

The foundation model screening study offers practical application information for the growing number of vision
foundation models. Frozen utility for this (out of domain) task was largely insensitive to the pretraining
objective: contrastive, self-distillation, masked-reconstruction, and segmentation backbones all
performed comparably. The clearest failure was a backbone whose augmentation-invariance objective
is trained to discard the intesity and appearance variation indicating artifact
severity (the SimCLR-pretrained BrainIAC). Among the distillation teachers, a pathology-pretrained model may have contributed most improvement to the student, which we attribute to the fine-grained textural cues
that some image artifacts share with histopathology imaging.

Several limitations of this study should be noted. First, the LISA data are
magnitude-only and image-domain, so the phase information that most directly encodes motion- and zipper-type
corruption is unavailable. In the scenario of a point-of-care deployment of artifact detection alongside an operating system, complex k-space data would very likely be available and would provide much additional useful information to a classification model. Second, pooling
the external low-field corpus produced only a marginal, within-noise change in the overall
composite. The benefit of this modification was to redistribute model capacity toward the weak, $B_0$-driven classes,
most notably Distortion, while reducing cross-seed variance. A larger and more diverse exact-field
low-field corpus may be required to improve the overall score further. This study was further limited by the inclusion only of $T_2$-weighted images, in alignment with the challenge's available data. Performance across other contrasts or in mixed-contrast datasets has not been demonstrated and should be investigated. Third, the LISA data are drawn from three sites
with different overall image quality and individual artifact prevalences, and we observed that generalization across sites,
rather than model capacity, was a factor limiting performance. Addressing this
cross-site distribution shift~\cite{esteban2017mriqc} is a potential direction for future work. Finally, an evaluation of the distilled backbone across additional downstream
low-field MRI tasks, such as segmentation or image enhancement, remains to be performed.

\begin{credits}
\subsubsection{Data and Code Availability} Data used as a part of this study were provided by the LISA challenge organizers and cannot be shared by the authors. Upon publication, code will be made available at \url{https://github.com/jonbmartin/LoFi-RADIO-MICCAI}.
\subsubsection{\ackname}
Research reported in this publication was supported by the National Heart, Lung, and Blood
Institute of the National Institutes of Health under Award Number K25HL183904, the National Institute of Biomedical Imaging and Bioengineering of the National Institutes of Health under Award Number R01EB029443, and the American Association of University Women (AAUW) Award Number G-2025-14902. The content is
solely the responsibility of the authors and does not necessarily represent the official views
of the National Institutes of Health or AAUW.

\subsubsection{\discintname}
The authors have no competing interests to declare that are
relevant to the content of this article.
\end{credits}
%
% ---- Bibliography ----
%
% BibTeX users should specify bibliography style 'splncs04'.
% References will then be sorted and formatted in the correct style.
%
% \bibliographystyle{splncs04}
% \bibliography{mybibliography}
%

\end{document}

%% file: tab_screening.tex
\begin{table}[t]
\centering
\caption{Uniform frozen-feature screen of candidate backbones for ULF neonatal artifact grading. Every row is measured at a single-arm probe with the tuned head, on the same subject-grouped folds and 10 seeds, compared to DINOv2 (the anchor). $\overline{|\cos|}$ and eff-rank are reported as a simple potential degeneracy screen.}
\label{tab:screening}
\setlength{\tabcolsep}{4pt}
\small
\resizebox{\textwidth}{!}{%
\begin{tabular}{lllccr@{\,$\pm$\,}lr}
\toprule
Backbone & Objective & Pretrain Domain & $\overline{|\cos|}$ & eff-rk & \multicolumn{2}{c}{Composite} & $\Delta$ \\
\midrule
\multicolumn{8}{l}{\textit{Distilled / gated teachers (kept)}} \\
\textbf{Phikon} & iBOT & pathology & 0.81 & 0.010 & 0.8287 & 0.0027 & +0.0038$^\ast$ \\
DINOv2 & self-distill. & natural & 0.87 & 0.029 & 0.8249 & 0.0021 & --- \\
BiomedCLIP & contrastive & biomedical & 0.92 & 0.019 & 0.8240 & 0.0031 & -0.0009 \\
SAM & segmentation & natural & 0.99 & 0.017 & 0.8158 & 0.0043 & -0.0091$^\ast$ \\
\midrule
\multicolumn{8}{l}{\textit{Other screened candidates}} \\
RadioDINO & DINO & radiology & 0.69 & 0.033 & 0.8246 & 0.0023 & -0.0003 \\
MAE & masked recon. & natural & 1.00 & 0.009 & 0.8213 & 0.0021 & -0.0036$^\ast$ \\
SAM2 & segmentation & natural & 0.90 & 0.003 & 0.8212 & 0.0024 & -0.0037$^\ast$ \\
Triad & masked recon. (3D) & brain MRI & 1.00 & 0.002 & 0.8121 & 0.0031 & -0.0128$^\ast$ \\
ARNIQA & IQA contrastive & natural & 0.90 & 0.002 & 0.8118 & 0.0033 & -0.0131$^\ast$ \\
BrainIAC & SimCLR (3D) & brain MRI & 0.98 & 0.003 & 0.7633 & 0.0035 & -0.0616$^\ast$ \\
\bottomrule
\end{tabular}%
}
\\[2pt]
{\footnotesize $^\ast$paired $|t|\ge2.5$ and $|\Delta|>0.002$.}
\end{table}